# Time Series Modeling for Dream Team in Fantasy Premier League

Gupta, Akhil[1]

[1] Department of Mechanical and Industrial Engineering
Indian Institute of Technology, Roorkee, India

**Abstract -** The performance of football players in English Premier League varies largely from season to season and for different teams. It is evident that a method capable of forecasting and analysing the future of these players' on-field antics shall assist the management to great extent. In a simulated environment like Fantasy Premier League, enthusiasts from all over the world participate and manage the players catalogue for the entire season. Due to the dynamic nature of points system, there is no known approach for the formulation of dream team. This study aims to tackle this problem by using a hybrid of Autoregressive Integrated Moving Average (ARIMA) and Recurrent Neural Networks (RNNs) for time series prediction of player points and subsequent maximisation of total points using Linear Programming (LPP). Given the player points for the past three seasons, the predictions have been made for the current season by modelling differently for ARIMA and RNN, and then creating an ensemble of the same. Prior to that, proper data preprocessing techniques were deployed to enhance the efficacy of prepared model. Constraints on the types of players like goalkeepers, defenders, midfielders and forwards along with the total budget were effectively optimised using LPP approach. The validation of the proposed team was done with the performance in ongoing season, where the players outperformed as expected, and helped in strengthening the feasibility of the solution. Likewise, the proposed approach can be extended to English Premier League by official managers on-field.

## 1. Introduction

Since it's invention way back in the mid-nineteenth century, Football (as the Europeans call it) has continually gained popularity across the world. Being a team sport played between two teams of eleven players with a spherical ball, it has more than 250 million followers in over 200 countries [1]. In the modern era, the game has reached a stage where any activity of every player is under supervision at all times. Consequently, the prediction of their performances is of utmost significance when it comes to the success of individual teams. There exist different measures of performance for footballers, namely, goals for forwards, assists for midfielders, saves for goalkeepers or standardised points system. This paper deals with the aforementioned idea of forecasting the future antics of a player using the points attained by him in previous encounters based on his on-field performance.

### 1.1. *Fantasy Premier League*

English Premier League (EPL) is an English professional league for men's association football clubs [2]. It is contested by 20 clubs and operates on a system of promotion and relegation. Owing to its eminence in sports arena as the most-watched sports league in the world, it generates high revenue and has therefore, garnered broad attention from prominent industrialists and businessmen. Fantasy Premier League (FPL), a simulated version of EPL, is a chance for football enthusiasts to create their own squad of 15 players and battle across the entire world to atop the leaderboard. Organised by EA Sports, FPL aims at providing lucrative prizes and heaps of appreciation to the winners. The study is focussed at helping a novice aficionado to obtain a head-start when contesting FPL, and exhibit an approach better than the traditional naive ones.

---

[1] Gupta, Akhil. Tel.: +91-9760266638
 *E-mail address:* akhilg.iitr@gmail.com

1.2.   *Dream Team Formulation*

As indicated in 1.1, the main objective of FPL is to maintain a team of 15 players whose performance will be the sole determiner of the leaderboard position. Ideally, the squad is formulated after proper analysis of footballers across all 20 participating clubs. Given a fixed budget of £100 Million, the team must consist of 2 Goalkeepers, 5 Defenders, 5 Midfielders and 3 Strikers. Additionally, it should generate the highest team score possible, which in turn, depends on the game-weeks in which at least one player has stepped on the field.

Using the points for the past years, time series forecasting techniques were leveraged to predict next year's performance. To be robust in approach and consider both linearity and non-linearity, Autoregressive Integrated Moving Average (ARIMA) and Recurrent Neural Networks (RNNs) were considered for modelling. After obtaining the points for the entire roster, Linear Programming (LPP) basics were applied to adhere to all constraints and finalise a suitable starting dream team.

1.3.   *Sports Analytics*

Relevant and historical statistics, if properly used, can provide competitive advantage to a team or an individual. Coming into the forefront after the release of 2011 film, *Moneyball* [3], where analytics was used to build a competitive team on a minimal budget, this field has never looked back. This paper covers both on-field and off-field aspects, as it seeks to move from performance analysis to managerial decisions.

One of the ill-effects though has been its impact on sport gambling. It has escalated to another level as betters now have more information than ever at their disposal to aid in decision making. Being inevitable as such, increased companies and startups in this domain are not helping the cause at all.

2.   **Related Work**

When it comes to Fantasy Football, there is no known published work which covers the aspects of team formulation or point-based analytics. There have been some researchers who have talked about the performance of referees and the effect of age [4-5]. Paolo Cintia is a researcher in Sports Science and deliberates upon the harsh rule of goals [6] and evaluates the performance of football teams using a network-based approach [7]. Significant research on player hamstring history [8] and high running during football games [9] has helped in better management of the bench by football managers.

The only aspect of this study that has seen some prior research is Performance Analysis (PA). Craig Wright [10] talks about its application in the football coaching process, whereas, Isidoro [11] measures the productivity and efficiency of the football teams in EPL. This paper aims to create an everlasting remark in the field of football analytics as it introduces the highly accurate neural networks approach.

Over the past several decades, much effort has been devoted to the development and improvement of time series forecasting models (Zhang [12]). Traditional statistical models including moving average, exponential smoothing, and ARIMA are linear in that predictions of the future values are constrained to be linear functions of past observations. Because of their relative simplicity in understanding and implementation, linear models have been the main research focuses and applied tools during the past few decades. To overcome the restriction of the linear models and to account for certain nonlinear patterns observed in real problems, several classes of nonlinear models have been proposed in the literature. These include the bilinear model [13], the threshold autoregressive (TAR) model [14], and the autoregressive conditional heteroscedastic (ARCH) model [15]. Although some improvement has been noticed with these nonlinear models, the gain of using them to general forecasting problems is limited [16]. Because these models are developed for specific nonlinear patterns, they are not capable of modeling other types of nonlinearity in time series. More recently, Artificial Neural Networks (ANNs) have been suggested as an alternative due to their flexible nonlinear modeling capability. However, a totally unexplored category of neural networks called the Long Short-Term Memory Recurrent Neural Networks (LSTM-RNN) is the main focus of research here, as it is combined with ARIMA which captures linearity with a great efficiency.

After time series forecasting, it is imperative to focus on utilisation of linear programming to get the best out of everything. Jahangir and Mehrzad [17] have evaluated the performance of Iranian football teams using Data Envelopment Analysis, which is a linear programming-based technique. Apart from that, there have been significant contribution across different spheres but lesser in sports arena. To keep it simple, this study uses the novel optimisation approach which takes in some constraints and maximises the objective function by Simplex method. To decrease the computational time, the solver was appropriately chosen.

This paper takes into account all of the prior contributions by knowledgeable researchers and aims to move forward by delivering concrete results in sports performance analytics. More of original concepts and ideas have been proposed and presented. This should open opportunities for research into individual player tracking using other factors like demographics, food habits, training time, sleep time, medical health status, etc.

3. **Data**

The data utilised in the research was collected from the official Fantasy Premier League website [18] which has all the details about various players taking part in the English Premier League for the past five to six years.

Besides official FPL, few *Kaggle* kernels [19] were used to cross-check the historical data points to avoid any loss or misinterpretation of data. There were two kinds of datasets that were fetched using Data Scraping techniques in *Python 3*. Master FPL dataset consisted of mapping of players onto their position, team and cost. The shape of this dataset was estimated to be (Number of Players i.e. 584, 5). A sample of the same can be seen in Table 1.

Table 1. Master FPL Dataset

| First Name | Surname | Position | Team | Cost |
|---|---|---|---|---|
| Almen | Abdi | MID | WAT | 5000000 |
| Benik | Afobe | FWD | BOU | 5700000 |
| Toby | Alderweireld | DEF | TOT | 6500000 |

For time series forecasting, historical data was of as much importance. This was extracted year-wise for all gameweeks that each player played. For the following study, the data was used for 3 seasons, namely, 2013-14, 2014-15 and 2015-16. This yearly dataset listed player-wise sorted data along with gameweek, player ID, team, weekly points and running sum of points (Table 2).

Table 2. Points Dataset for FPL Season 2014-15

| Name | Gameweek No. | PID | Team | Weekly Points | Running Sum of Points |
|---|---|---|---|---|---|
| Zverotic | 12 | 562 | FUL | 1 | 1 |
| Zverotic | 13 | 562 | FUL | 1 | 2 |
| Zverotic | 19 | 562 | FUL | -1 | 1 |
| Stones | 24 | 524 | EVE | 2 | 16 |
| Stones | 28 | 524 | EVE | 8 | 24 |

The above table shows records in gameweek sequence for two random players *Zverotic* and *Stones*. The gameweeks missing in between indicate player's absence from those matches. PID has been provided for correct identification of the same player who has played for different clubs across various seasons. Running sum of points indicate the cumulative points in that particular season. Using the data for three seasons, the aim is to forecast the points for next season 2016-17 and formulate the dream team by taking different constraints into consideration.

4. **Proposed Methodology**

The entire process of generating the dream team can be broadly divided into three phases as shown below (Fig. 1):

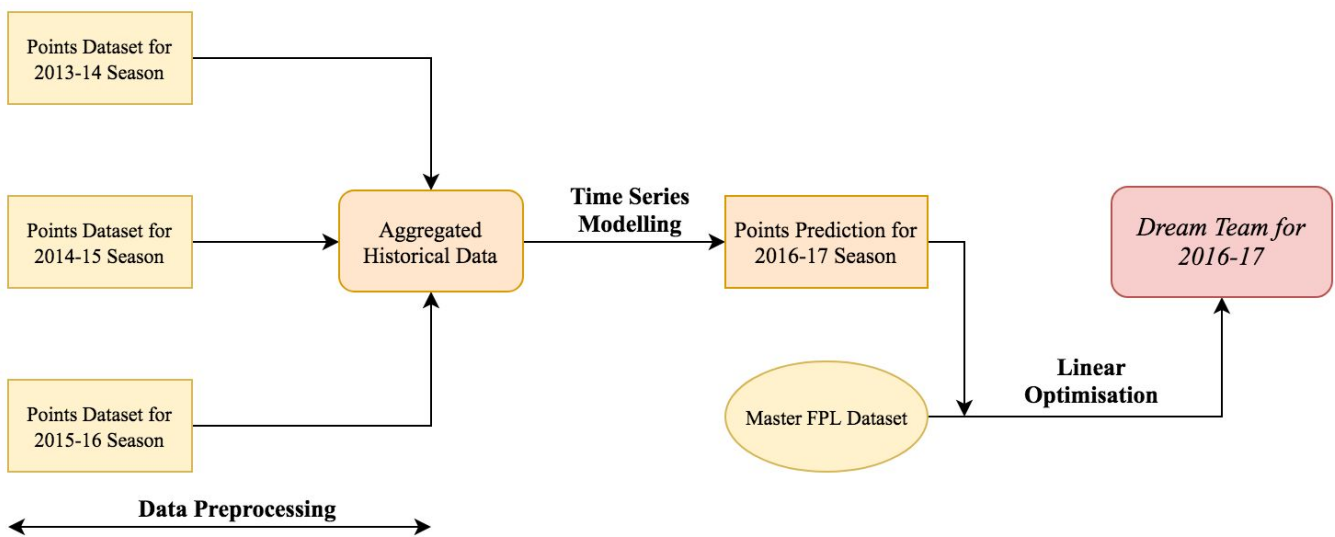

Fig. 1. Entire procedure for formulation of the ideal squad of 15

### 4.1. Data Preprocessing

As mentioned by Han [20], today's real-world databases are highly susceptible to noisy, missing, and inconsistent data due to their typically huge size (often several gigabytes or more) and their likely origin from multiple, heterogeneous sources. Low-quality data will lead to low-quality mining results. The appropriate principles, suiting the data format and type of data, were applied for data preprocessing.

- First of all, the missing values were identified in the *Name* field and replaced by '-' as it was a string literal.
- In the Master FPL dataset, the cost was scaled down to in terms of millions by dividing by 1000000.
- New blank *ID* field was created for matching of the two datasets for analysis. (Table 3)
- Points data from all seasons was merged into a single data-frame in *Python 2.7*. Columns for all gameweeks were created, whether or not the player played. If he didn't play, zero score was assigned for that week. This ensured the consistency of rows between Points and Master FPL dataset.

The next important step was to populate the *ID* field to ensure that correct player's performance is being tracked. For the same, team-wise examination was performed. Those players were removed who had joined EPL this season as their historical data wouldn't be available. Also, the dormant footballers, i.e., the ones who had left the league sometime in the past three seasons were cleaned up. This eliminated around 44% players from the entire roster.

Chronologically arranged, the points dataset had 114 values of points for a single player (38 for each season). These shall be treated as separate time series for each player, and future modeling be done accordingly.

Table 3. Modified Player Data having ID 236.

| First Name | Surname | Position | Team | Cost | ID |
|---|---|---|---|---|---|
| Daniel | Sturridge | FWD | LIV | 9.0 | 236 |

Table 4. Sample of Points Data for ID 236.

| Name | Team | ID | GW1 | GW2 | GW3 | GW4 | GW5 |
|---|---|---|---|---|---|---|---|
| Sturridge | LIV | 236 | 3 | 4 | 1 | 2 | 0 |

Table 4 shows a slice of 5 week values from a total of 114 distinct columns for *Daniel Sturridge*. The above tables demonstrate the preprocessing done for better predictions.

### 4.2. Time Series Modeling

This section uses historical values and associated patterns to predict the future activity. For improved accuracy of predictions, two different models were tried separately and their results were ensembled to come up with a hybrid version:

- *ARIMA*, or Autoregressive Integrated Moving Average
- *LSTM-RNN*, or Long Short-Term Memory Recurrent Neural Networks

Each one is able to cover up for the shortcomings of the other one. ARIMA's major limitation is the pre-assumed linear form of the model. LSTM ensures that the nonlinearity is captured to a large extent, and the model is not biased entirely towards linear components. Meanwhile, LSTM relies heavily on persistence and thus, arises the need for huge datasets for lesser error of forecast. As 114 values are not too huge, it may exhibit some deviations from perfection. ARIMA barges in here to make sure that these deviations are not that significant.

Time series data of each player (326 in total) was used for predictions for the next one year, i.e. next 38 values. ARIMA and LSTM-RNN were executed separately and then their results mixed in some proportion (p1 and p2) which shall be explained ahead. These 38 gameweek points were added then to find the total points for every player in the upcoming 2016-17 season.

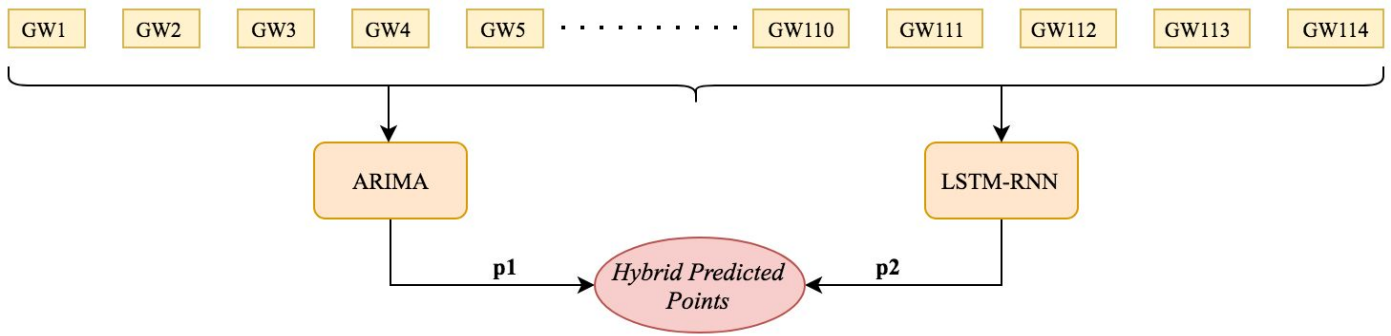

Fig. 2. Preparation of Ensembled points using Historical Gameweek values

### 4.3. *Linear Optimisation*

Commonly known as Linear Programming (LP), this seeks to achieve the best outcome in a mathematical model whose requirements are represented by linear relationships. An objective function is maximised or minimised, with respect to certain constraints which vary from situation to situation.

Once the predicted points for all 326 players were obtained as explained in 4.2, there was a need to maximise the total points for the desired dream team of 15 players. The *Pulp* library of Python 2.7 was used for applying the concepts of LPP. Formulation of the problem was done as shown in (1). Notations used are as follows:

- $z_i$, the forecasted points of the $i^{th}$ player for 2016-17 season
- $c_i$, the cost of the $i^{th}$ player
- $y_i$, binary value indicating whether the $i^{th}$ player is included in the dream team or not
- $n$, the total number of players concerned
- $F$, the set of players who play in Forward (FWD) position
- $M$, the set of players who play in Midfielder (MID) position
- $D$, the set of players who play in Defender (DEF position
- $G$, the set of players who play in Goalkeeper (GLK) position

$$Max. \sum_{i=1}^{n} z_i * y_i$$

subject to constraints:

$$\sum_{i=1}^{n} c_i * y_i <= 100.0 \quad (1)$$

$$\sum_{i \in F} y_i = 3 \qquad \sum_{i \in M} y_i = 5$$

$$\sum_{i \in D} y_i = 5 \qquad \sum_{i \in G} y_i = 2$$

$$y_i \in \{0, 1\}; \; 1 <= i <= n$$

The solution of this LPP is given in the form of certain values of *i*, which are then mapped onto the Master FPL dataset to obtain the exact list of players selected for dream team 2016-17. Due to less data for optimisation, the Pulp solver didn't take long to provide the final 15.

This approach of optimisation helped a great deal in reducing the effort after time series forecasting was done. Proper understanding of the problem is necessary for formulation of objective function and constraints. Once done, it is a matter of input-output.

## 5. **Experimental Results**

This section focuses on showcasing the results obtained after running time series models on FPL dataset. Owing to large number of time series generated, the process of prediction was automated after setting optimal parameters. LSTM-RNN works on a large architecture and explaining the entire process is out of the scope of this study. However, for fitting ARIMA model to the points time series data, three-stage procedure of model identification, estimation of model parameters and diagnostic checking of estimated parameters was employed. To determine the possible persistence, the autocorrelation function (ACF) and the partial autocorrelation function (PACF) were used. Using Akaike Information Criteria (AIC), the best fitted model has been identified out of the various competing models. For demonstration purpose, the seasonal and nonseasonal components of 5 random players are indicated in Table 5. Generally, lower AIC value indicates the presence of a better fit.

Table 5. Summary of statistical parameters of the best fitted ARIMA models

| Name | Team | Pos | ID | Model | AIC |
| --- | --- | --- | --- | --- | --- |
| de Gea | MUN | GLK | 26 | $(1.1.1)(0.0.1)_{12}$ | 121.3 |
| Costa | CHE | FWD | 145 | $(1.0.1)(0.1.1)_{12}$ | 91.4 |
| Can | LIV | MID | 321 | $(1.1.0)(1.0.0)_{12}$ | 140.1 |
| Fabian | MCI | MID | 84 | $(1.0.1)(1.0.0)_{12}$ | 100.5 |
| Debuchy | ARS | DEF | 210 | $(1.1.1)(0.1.1)_{12}$ | 96.1 |

After fitting ARIMA to capture linear components, LSTM-RNN was executed on the entire roster. Both the models were validated using the 2015-16 season by training on 2013-14 and 2014-15. To depict the accuracy and robustness of the hybrid model, two kinds of players have been graphically analysed. *Jamie Vardy* (Fig. 3) has played almost all matches in the last season, whereas, *Bacary Sagna* (Fig. 4) is prone to injuries and benchwarmer for the majority period. (Validation values only)

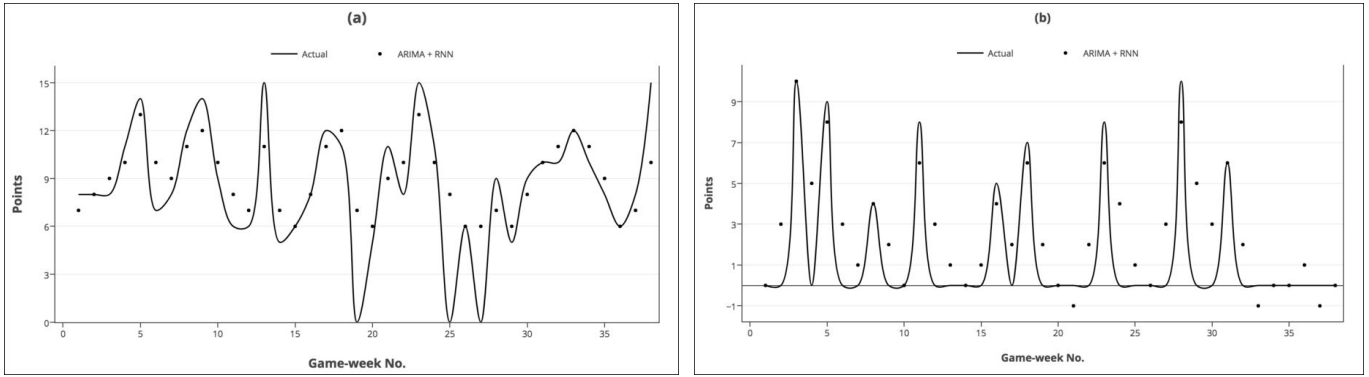

Fig. 3. Comparison of Actual and Predicted values for (a) *Jamie Vardy* (b) *Bacary Sagna*

For mathematical calculation of deviation from the actual values, error metric *Root Mean Squared Error* (or RMSE) [21] was put to use. This was further utilised in finding the proportion of ARIMA (**p1**) and LSTM-RNN (**p2**) in the hybridised version.

To keep the ensemble simple, the step size of proportion increment/decrement was chosen to be 10%. Starting with *pure ARIMA*, the aim was to stop at *pure LSTM-RNN* and identify the least error obtained in the process.

Table 6. Determination of optimal ensemble proportions

| % ARIMA (p1) | % LSTM-RNN (p2) | RMSE (*Vardy*) | RMSE (*Sagna*) |
|---|---|---|---|
| 100 | 0 | 3.412 | 4.156 |
| 90 | 10 | 3.304 | 3.588 |
| 80 | 20 | 2.965 | 3.405 |
| 70 | 30 | 2.693 | 3.194 |
| 60 | 40 | **2.539** | 2.747 |
| 50 | 50 | 2.591 | 2.798 |
| 40 | 60 | 2.808 | 2.310 |
| 30 | 70 | 3.005 | **2.013** |
| 20 | 80 | 3.296 | 2.555 |
| 10 | 90 | 3.114 | 2.894 |
| 0 | 100 | 3.532 | 2.639 |

In this way, the optimal coefficients for the two models were determined for every player. Reduction in RMSE directly meant better predictions.

Keeping future formulation of dream team in perspective, a common *(p1,p2)* was identified which would significantly reduce the computational time for a new player and ensure trade-off between time and accuracy (Fig. 4). For the preprocessed master FPL dataset, the final proportions came out to be **(0.4,0.6)** after rounding off to the nearest tenths.

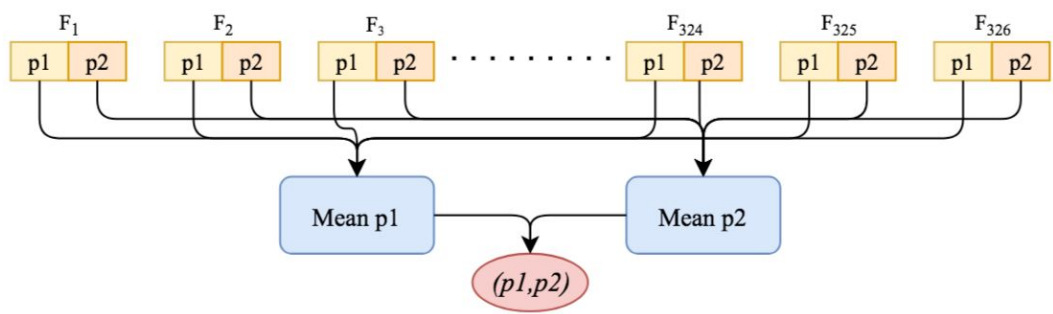

Fig. 4. Generalisation of hybrid model by finding mean coefficients

6. **Discussions and Intelligence Gained**

After successful time-series forecasting, linear optimisation was used to obtain the dream team. The squad of fifteen has been deliberated upon in this section.

Solution to (1) gave the maximum possible score within the stipulated budget which came out to be *3718 points* (Table 7). For better understanding of the performing players, some new features were computed by assuming that the player shall play almost the same matches as he played in 2015-16 season. These were:

- *PPM* (Points per match) - Average points a player is expected to score in a match. The higher, the better.
- *CPP* (Cost per point) - Helped in determining most valuable players, i.e. cheaper and more points. The lower, the better.
- *CPI* (Cost-point index) - A combination of PPM and CPP, this was calculated for those players expected to play more than 20 matches. PPM and CPP were scaled down to lie between 0 and 1, and then added by giving equal weightage to each. This weightage can be varied for the desirable quality in the team.

Table 7. Players in Dream Team 2016-17

| Name | Team | Position | Points | Matches | PPM | CPP | CPI |
|---|---|---|---|---|---|---|---|
| Branislav Ivanovic | CHE | DEF | 302 | 37 | 8.162 | 0.020 | 0.834 |
| Cesc Fabregas | CHE | MID | 249 | 33 | 7.545 | 0.029 | 0.760 |
| Chris Brunt | WBA | DEF | 236 | 34 | 6.941 | 0.021 | 0.765 |
| Chris Smalling | MUN | DEF | 223 | 25 | 8.920 | 0.026 | 0.843 |
| Eden Hazard | CHE | MID | 275 | 37 | 7.432 | 0.037 | 0.715 |
| Jamie Vardy | LEI | FWD | 293 | 35 | 8.371 | 0.034 | 0.779 |
| Joe Hart | MCI | GLK | 194 | 36 | 5.389 | 0.027 | 0.652 |
| Jose Fonte | SOU | DEF | 278 | 37 | 7.514 | 0.019 | 0.802 |
| Romelu Lukaku | EVE | FWD | 301 | 36 | 8.361 | 0.030 | 0.792 |
| Ross Barkley | EVE | MID | 249 | 29 | 8.586 | 0.031 | 0.806 |
| Saido Berahino | WBA | FWD | 222 | 38 | 5.842 | 0.028 | 0.671 |
| Simon Mignolet | LIV | GLK | 176 | 35 | 5.029 | 0.028 | 0.628 |
| Stewart Downing | MID | MID | 200 | 37 | 5.405 | 0.027 | 0.655 |
| Toby Alderweireld | TOT | DEF | 293 | 26 | 11.269 | 0.022 | 0.987 |
| Tom Cleverley | EVE | MID | 227 | 32 | 7.094 | 0.023 | 0.761 |

It is quite evident that this team is full of high performers and some cost-efficient players. LPP approach intuitively helps in achieving the trade-off. Apart from the dream team, there were some players with exceptional CPI worth-mentioning (Table 8).

Table 8. Key players in 2016-17 (Using CPI)

| Goalkeepers | Defenders | Midfielders | Forwards |
|---|---|---|---|
| Pantilimon (0.66/WAT) | John Stones (0.85/MCI) | Riyad Mahrez (0.76/LEI) | Sergio Aguero (0.77/MCI) |
| Thibaut Courtois (0.62/CHE) | Kolarov (0.75/MCI) | Oscar (0.76/CHE) | Costa (0.75/CHE) |
| Lukasz Fabianski (0.62/SWA) | Laurent Koscielny (0.71/ARS) | Alexis Sanchez (0.75/ARS) | Olivier Giroud (0.72/ARS) |

This study helped in better understanding of performances in EPL, and following intelligence was gained:

- Loyalty was discovered as a major factor in enhancement of a player's performance. For the dream team, average stay period came out to be 5 years, clearly indicating that the more you play for a particular team, the better your chances of excelling.
- As expected, most of the players were natives of England, but surprisingly, the other big majority belonged to Belgium, expected to be a supplier of great young talent.
- On an average, Goalkeepers performed best on CPI scale followed by defenders. Maximum variation could be seen among the forwards as the sport witnesses mixed array of players in that position.
- Hull City performed poorly on CPI, whereas, Chelsea and Arsenal were the table toppers, producing match-winners in all positions from GLK to FWD.
- Team's gameplay can be estimated from the eleven that it fields on the pitch. Manchester City are unbeatable when it comes to defence, and Chelsea are unstoppable with their sublime list of attacking players.

For cross-validation and establishment of efficacy of the approach followed, this team was tracked for 2016-17 season and the RMSE between predicted points and actual points was calculated.

The hybrid performed with a reasonably good accuracy exhibiting an error of only **87 points** on the negative side for the entire team, i.e. over-forecasting. The precise inability of the model to pick out random absence from the playing team (sharp decrease to zero) was attributed as the reason for this. Model that assists in predictions for non-continuous time series may further help in reducing this inflated error. Some of the players in Dream team won Player-of-the-Month awards also, thereby, strengthening the feasibility of the proposed solution.

7.  **Conclusion and Future Work**

The increasing use of Analytics and Statistics in the field of Sports has intensified the global competition and paved way for remarkable development. Through this study, the aim has been to introduce the latest concepts of Machine Learning to find insights about the footballers playing in English Premier League. Powerful predicting nature of Long Short-Term Memory Recurrent Neural Networks (LSTM-RNN) has been leveraged to identify the non-linear patterns in points-based time series data of the players with historical data and supposed to play in the ongoing season 2016-17. As the amount of data increases, the accuracy of model shall improve, thereby establishing the persistence of RNN over time.

Fantasy Premier League (FPL), a simulated version of the actual Premier League has given common man the opportunity to don a Manager's hat and rally the troops as per his wishes. This formulation of dream team gives a unassailable head-start to any novice fanatic. It takes into account all constraints with the help of properly designed Linear Programming environment. Another unexplored domain is tracking of player's performance beforehand. In today's fiercely competitive world, it's imperative to ensure that highly efficient personnel are deployed at all times. This model shall aid the manager in figuring out the low points in a player's season, and accordingly making the necessary adjustments, either by pushing him or replacing him. Therefore, this study stands at the intersection of machine learning and sports analytics, and seeks to facilitate extensive and varied research in this field.

Future work can be expanded to other football leagues in the world for better feedback on the proposed model. As the study hinges mostly around working with time series data, it's usage in other sports like Runs in Cricket, Baskets in Basketball and Winners in Tennis can be explored and studied. More features about the player could be used in understanding his worth as an asset to the team. Features like fitness levels, eating habits, demographics, records against specific oppositions, etc. can be of great help. Fitness and commitment levels may assist in enhanced determination of the number of matches the player can be expected to play. Further research into models that take into account, sudden peaks or non-continuous time series data can be looked upon. Accurate forecasting of player's playing time shall always remain the key. Importantly, it should be ensured that this analysis doesn't end up in wrong hands and result in malpractices i.e. sports betting or gambling. The concerted goal of aggrandizing the management of sports activities across the globe with the help of Analytics and Mathematics should be gradually realised, leading to revamped viewership for the billions.

**Acknowledgements**

I would like to take this opportunity to thank my college-mate Nikita Saini for helping me with the fitting of ARIMA models leading to accurate prediction of points for the 2016-17 season. It involved fundamental statistical computations.

I would also like to express my gratitude to Dhish Kumar Saxena for sharing his knowledge of the subject *Product and Process Optimisation* and motivating me for pursuing this research.